\documentclass{emulateapj}
\usepackage{graphicx}

\newcommand{\lya}{Ly$\alpha$~}
\newcommand{\Lya}{Ly$\alpha$}

\shorttitle{The lack of §Lya in $z=7$ candidates}
\shortauthors{Fontana et al.}

\begin{document}


\title{The lack of intense Lyman~$\alpha$ in ultradeep spectra of
  $z=7$ candidates in GOODS-S: imprint of reionization?}


\author{A. Fontana\altaffilmark{1},
E. Vanzella \altaffilmark{2}
L. Pentericci, \altaffilmark{1}
M. Castellano \altaffilmark{1}
M. Giavalisco \altaffilmark{3}
A. Grazian \altaffilmark{1}
K. Boutsia \altaffilmark{1} 
S. Cristiani \altaffilmark{2} 
M. Dickinson\altaffilmark{4}
E. Giallongo \altaffilmark{1}
M. Maiolino, \altaffilmark{1} 
A. Moorwood \altaffilmark{5}
P. Santini \altaffilmark{1} }

\affil{INAF Osservatorio Astronomico di Roma,  Via Frascati 33,00040 Monteporzio (RM), Italy}
\affil{INAF Osservatorio Astronomico di Trieste, Via G.B.Tiepolo 11, 34131 Trieste, Italy}
\affil{Department of Astronomy, University of Massachusetts, 710 North Pleasant Street, Amherst, MA 01003 }
\affil{National Optical Astronomy Observatory, PO Box 26732, Tucson, AZ 85726, USA}
\affil{European Southern Observatory, Karl-Schwarzschild Strasse, 85748 Garching bei Munchen, Germany}
\email{adriano.fontana@oa-roma.inaf.it}





\begin{abstract}
  We present ultradeep optical spectroscopy obtained with FORS2 on VLT
  of seven Lyman--break galaxy (LBG) candidates at $z>6.5$ selected in
  the GOODS--S field from Hawk--I/VLT and WFC3/HST imaging. For one
  galaxy we detect a low significance emission line
  ($S/N\leq 7$), located at $\lambda=9691.5\pm 0.5$~\AA\  and with flux
  $3.4\times 10^{-18}$erg~cm$^{-2}$s$^{-1}$. If identified as \Lya, it places
  the LBG at redshift $z=6.972\pm 0.002$, with a rest--frame
  equivalent width  $EW_{rf}=13$\AA. Using Monte Carlo simulations
  and conservative EW distribution functions at $2<z<6$, we estimate
  that the probability of observing no galaxies in our data with
  $S/N>10$ is $\simeq 2$\%, and that of observing only one galaxy out of
  seven with $S/N=5$  is $\simeq 4$\%, but these can be as small
  as $\sim 10^{-3}$, depending on the details of the EW
  distribution. We conclude that either a significant fraction of the
  candidates is not at high redshift or that some physical mechanism
  quenches the \lya emission emerging from the galaxies at $z>6.5$,
  abruptly reversing the trend of the increasing fraction of strong
  emitters with increasing redshift observed up to $z\sim 6.5$. We
  discuss the possibility that an increasingly neutral intergalactic medium
  is responsible for such quenching.
\end{abstract}

\keywords{galaxies: distances and redshifts - galaxies: high-redshift - galaxies: formation}

\section{Introduction}

During the past year, a suite of new near--infrared (NIR) surveys has
extended the search for star--forming galaxies to redshift $6.5 \leq
z\leq 10$ using the well--proven Lyman--break technique \citep[][and
references therein]{Giavalisco2002}.  With respect to lower redshift,
the number density  of UV--selected galaxies decreases \citep[e.g.,][]{Ouchi2009, Mclure2009b,
  Castellano2010, Bouwens2009b, Wilkins2010}, their UV continuum
becomes bluer implying either reduced dust obscuration or poorer metal
enrichment or
both \citep[e.g.,][]{Finkelstein2009,Bouwens2010b,Schaerer2010}, and
their stellar masses are, on average, smaller than those of their
lower redshift counterparts \citep[e.g.,][]{Labbe2010}.  Unfortunately,
these results are based on color-selected samples with no
spectroscopic validation. 
At the time of writing,  spectroscopic detections of only a few individual
objects have been
obtained at $z>6.6$
\citep{Iye2006,Greiner2009,Salvaterra2009,Tanvir2009}.

The lack of knowledge of the true redshifts of the current
$z\sim 7$ candidates places significant limitations on our ability to robustly
measure the properties of the galaxies at this critical cosmic epoch. For
example, the fraction of interlopers and the redshift distribution of the
sample galaxies are necessary to robustly measure the UV luminosity 
function \citep[e.g.,][]{Reddy2009}. Currently, the former remains unknown, and
the latter is estimated with Monte Carlo simulations 
under various assumptions for the intrinsic distributions of UV spectral energy distribution (SED),
surface brightness and morphology, with the result that the measure of the luminosity function
remains subject to uncontrolled systematic errors.

In practice, given the marked decrease in sensitivity of current
spectroscopic observations at increasing redshift, the spectral
  confirmation of galaxies at $z>5$ relies heavily on their \lya
  emission line, \citep[][S10 and V09 in the following]{Stark2010,
    Vanzella2009}. Indeed, redshifts derived without \lya typically
  have lower confidence, although their number may be comparable
  \citep[][D10 in the following]{Douglas2010}.  The line in itself is
an important diagnostic of physical processes at work in the
galaxies \citep[e.g.,][]{Giavalisco1996,Pentericci2010,Shapley2003},
since its strength and velocity profile depend on the
instantaneous star-formation rate, dust content, metallicity, 
 kinematics and geometry of the interstellar medium. Particularly
relevant here is the evidence that the fraction of \lya emitters in
UV--selected samples increases with redshift \citep[V09,S10][S07 in
the following]{Reddy2009,Stanway2007} and that the fraction of
galaxies with a large \lya equivalent width (EW) is substantially
larger at fainter UV luminosities.

Finally, the very visibility of the \lya line during the ending phases
of the cosmic re--ionization is subject to the damping effect of an
increasing neutral intergalactic medium (IGM) \citep[e.g.,][]{Zheng2010,Dayal2010b}, expected
to attenuate most of its luminosity and make the earliest galaxies
consequently more difficult to identify. Hence, the line profile and
the evolution of its EW are sensitive diagnostics of the ionization
state of the high redshift IGM.

To address these issues we have started a campaign of spectroscopic follow-up
of $z\simeq7$ ``Z--dropout'' candidates, selected from high--quality 
imaging surveys obtained with VLT/Hawk-I and {\it HST}/WFC3.
In this paper we present the first results from a sample selected in the GOODS--S 
field \citep[][C10 in the following]{Castellano2010}.
All magnitudes are in the AB system, and we adopt
$H_0=70$~km/s/Mpc, $\Omega_M=0.3$ and $\Omega_{\Lambda}=0.7$.

\section{Targets and Spectroscopic Data}
This initial spectroscopic sample includes relatively bright Lyman break
galaxy (LBG)
candidates at $z>6.5$ (listed in Table \ref{targets}), five from the
Hawk--I images (4 from C10 and 1 from \citet{Hickey2009}) and two from
WFC3 \citep{Oesch2009b, Wilkins2009}, spanning the magnitude range
$Y\simeq 25.5-27.5$.  We filled empty slitlets in the multi-object slit masks with
other candidates of lower quality and/or at lower redshift, including
a candidate brown dwarf (\citet{Mannucci2007}, C10) and $i$--dropouts
from the GOODS survey not observed by V09.

\subsection{Observations}
Observations were taken in service mode with the FORS2 spectrograph on
the ESO Very Large Telescope, between 12 November 2009 and 14 January
2010.  We used the 600Z holographic grating, that provides the highest
sensitivity in the range $8000-10000$\AA\  with a spectral resolution
$R\simeq 1390$ and a sampling of 1.6\AA\  per pixel for a 1'' slit. The
data presented here come from the coaddition of 75 spectra of 842
seconds of integration each, on a single mask, for a grand total of
63150 s (17.5 hr), with median seeing around 0.8''. Each slitlet was
1'' wide and 14'' long, to maximize the number of slits
available while allowing a careful sky subtraction.
All our high priority targets were placed at the center of the slits,
and spectra were taken in series of three different positions, offset
by $\pm 2"$ in the direction perpendicular to the dispersion.

Since our objects are extremely faint, the slit centering was based on
the astrometry solution obtained from the Hawk--I images, which is
which is well aligned to the ACS one. We have directly verified this
by placing a few bright objects from the ACS catalogs in small slits,
and ensuring that they were correctly aligned during the
observations. It is also reassuring to note that three faint
  $i$-dropouts selected from the ACS catalog which were placed in
  slitlets using the same astrometry, have a clear \lya detection at
  $z \simeq 5.94$ (full details will be given in a future paper).

Data reduction was performed using an optimized version of the
recipes adopted in V09 and previous papers. After
standard flat-fielding and wavelength calibration, we subtracted
the sky emission lines with two different procedures. In the first
case (Polyn in the following) we fit polynomials of order $n$ (from 1
to 5) to
the sky intensity at each pixel position. 
This procedure in principle ensures the highest S/N, but is
sensitive to systematics induced by defects in the detector or in the
slit manufacturing. A safer but somewhat noisier approach
(ABBA in the following) is to subtract the sky background
between two consecutive exposures, exploiting the fact that the target
spectrum is offset due to dithering.  We found the spectra obtained
with the two techniques entirely consistent.  
Finally, spectra were flux-calibrated using the observations of
spectrophotometric standards. Slit losses are small, given the
extremely compact size of the targets and we neglect them in the
subsequent analysis.

\begin{figure}
 \epsscale{1.2}
\plotone{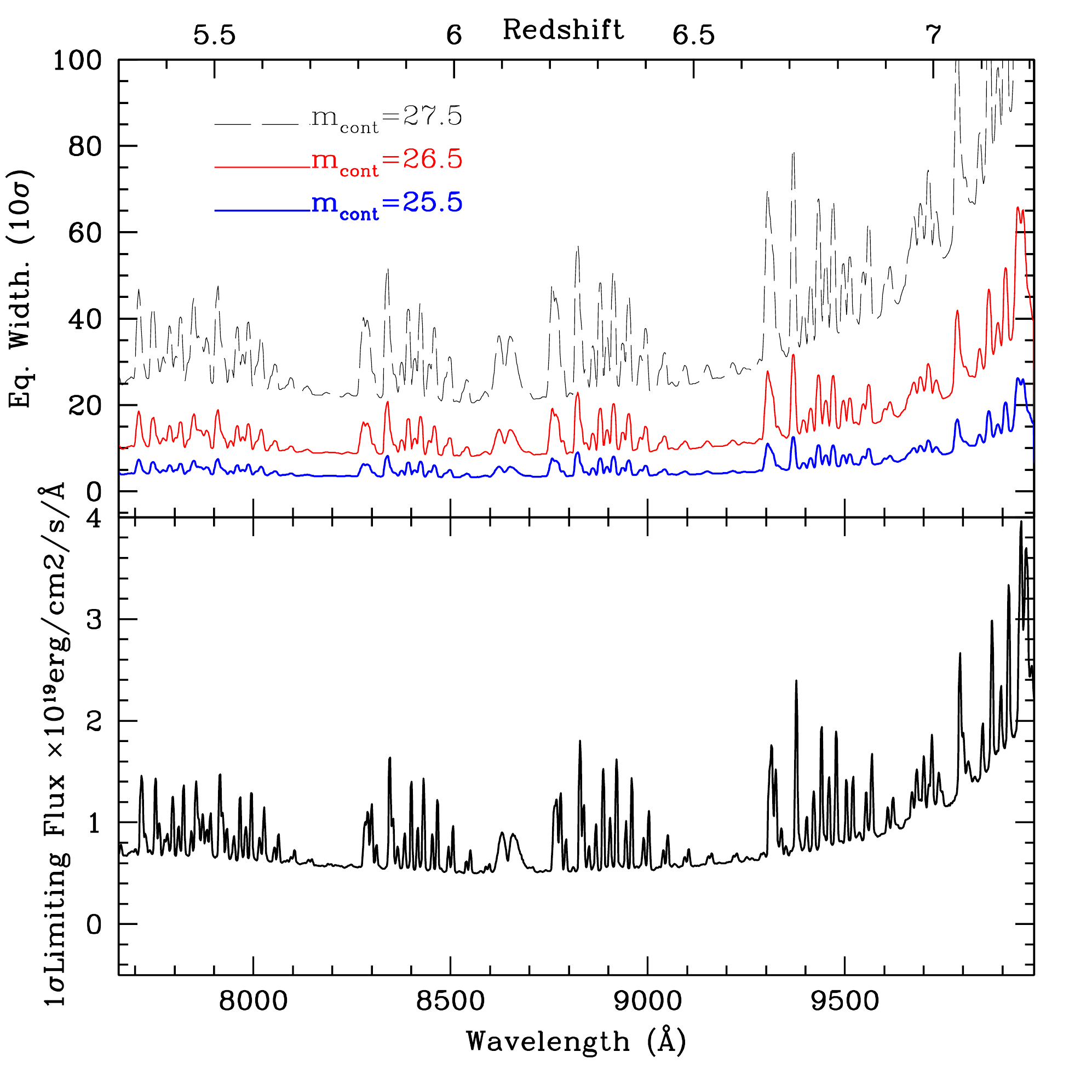}
\caption{ {\it Lower:} Limiting flux density (at $1\sigma$ level) resulting
   from our observations.  {\it Upper:} Corresponding $10\sigma$ limit on
   the rest-frame equivalent width of a \lya emission line as a function of
   redshift. Colors and line widths correspond to different observed
   magnitudes in the $Y$ band, as shown in the
   legend.}\label{EW_lambda}
 \end{figure}

The r.m.s.\ of the resulting spectra, which will be used later to
determine the probability of our results with a Monte Carlo
simulation, has been estimated ``by first principles'', i.e.,
computing the absolute r.m.s.\ of each frame from its raw counts $c$ as
$\sqrt{(c/g)}$ (where $g$ is the $e^-$-ADU conversion factor) and
propagating it through all the reduction steps. It turned to be in
excellent agreement with the observed r.m.s.\ in the region between
$8150$ and $8250$\AA\ that is devoid of sky lines and with the
predicted efficiencies estimated by the ESO exposure time calculator.
The resulting 1$\sigma$ limiting flux density is shown in the lower
panel of Fig.\ref{EW_lambda}.

\begin{table}
\caption{$z$-drop targets in GOODS-S}
\label{targets}
\centering
\begin{tabular}{ccccccc}
\hline 
ID & R.A. (deg) & DEC. (deg)&  Y & Z-Y & $M_{UV}^{(a)}$\\
\hline
G2\_1408$^b$& 53.177382& -27.782416& 26.37 & $>$2.1& -20.49\\
G2\_2370$^b$&53.094421 & -27.716847& 25.56 &1.68 & -21.27\\
G2\_4034$^b$& 53.150019& -27.744914& 26.35& $>$2.1& -20.50\\
G2\_6173$^b$& 53.123074& -27.701256& 26.53& $>$1.9& -20.33\\
H\_9136$^c$& 53.072574& -27.728610&  25.90& 1.29 & -20.94\\
W\_6$^d$&  53.100392 & -27.703847&   26.93& 1.17 & -20.38\\
O\_5$^e$&  53.177376 & -27.7920551&   27.52& 1.61 & -19.67\\
\hline
\end{tabular}
\\
\smallskip
\begin{tabular}{l}
a - Computed at $z=6.8$\\
b - \citet{Castellano2010}, $Y_{OPEN}$ Hawk-I\\
c - \citet{Hickey2009}, $Y_{OPEN}$ Hawk-I\\
d - \citet{Wilkins2009},  $Y_{098}$ WFC3 - ERS\\
e - \citet{Oesch2009b}, $Y_{105}$ WFC3 - HUDF\\
\end{tabular}
\\
\end{table}

 To obtain the corresponding limit on the detectable EW for a \lya
 line, we have computed three different cases, assuming continuum
 magnitudes of $m=25.5, 26.5, 27.5$, to span the luminosity range of
 our targets. For the computation we assume that the flux profile is a
 Gaussian with FWHM$=10$\AA. The resulting limiting EW is shown in the
 upper panel of Fig.\ref{EW_lambda}, computed at the 10$\sigma$ level.
 We could detect weak ($EW\simeq 5$\AA) \lya lines in our brightest
 galaxies, and even for the faintest ones we are able to reach $EW
 \simeq 50$\AA\ over a significant fraction of the redshift
 interval. This range of sensitivity is similar to that of $z\simeq
 5-6$ surveys (S07, V09, D10, S10).

 \begin{figure}
 \epsscale{1.2}
 \plotone{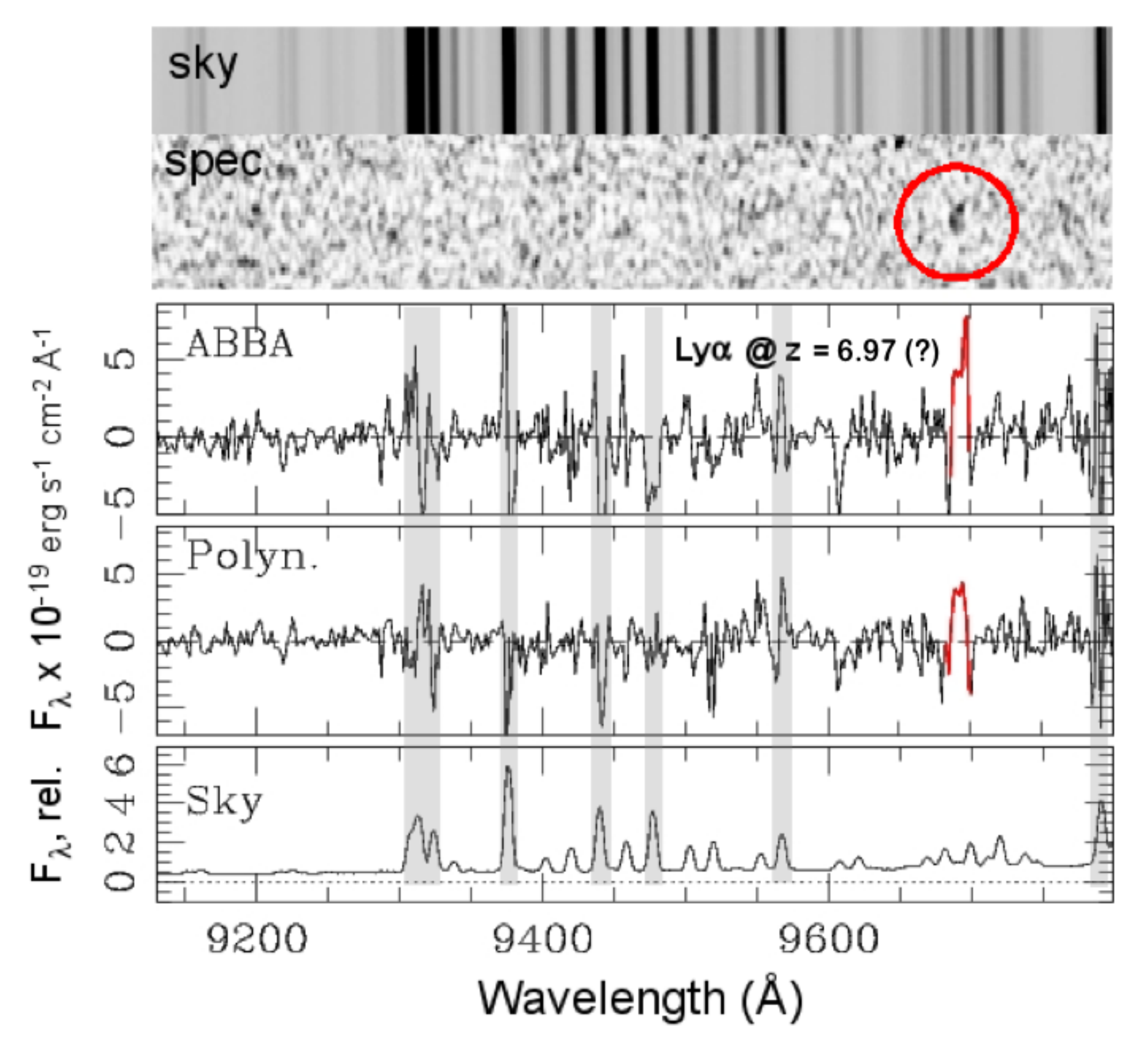}
 \caption{ Spectrum of the candidate G2\_1408, showing a tentative
   emission line at 9691.5\AA. The two upper panels show the 2--D
   spectrum of the sky emission and of the sky-subtracted object, as
   indicated. The x-axis is in wavelength, in the same range
   of the three spectra below. The 2-D spectrum of the galaxy has been
   divided by the r.m.s.\ to remove obvious spikes due to bright sky
   lines, and slightly filtered with an adaptive mesh.  The three 1-D
   spectra in the lower part show the extracted spectrum (over 4
   pixels) with the two different techniques for sky subtraction, and
   the sky emission at the same wavelengths (see legend). In these panels
   the spectra have not been divided by the r.m.s., nor filtered.}
\label{spectrum}
 \end{figure}

\subsection{Results}
We detect only one weak emission line, centered at $9691.5 \pm
0.5$\AA\  in the spectrum of the object G2\_1408. This galaxy is the
brightest candidate identified in the Hubble Ultradeep Field (HUDF) area, and one of the
brightest in C10.  It was first detected by \citet{Bouwens2004} in the
NICMOS HUDF data, and subsequently identified also by C10 and in the
HUDF WFC3 data
\citep{Bouwens2009b,Oesch2009b,Mclure2009b,Bunker2009}. From the clear
elongation observed in the WFC3 images, one can exclude the
possibility that it is a brown dwarf.  The 2-D and 1-D spectra of
G2\_1408 are shown in Fig.\ref{spectrum} The spectral feature is
extended over 4 pixels in the spatial direction, consistent with the
average seeing. The FWHM is $\simeq 10$\AA, significantly larger
than any feature due to noise.  The weak emission line has a total
observed flux of $ 3.4 \times 10^{-18}$erg~cm$^{-2}$s$^{-1}$. The formal S/N
is 7, but this estimate does not include systematic errors, and should
be considered as an upper limit.  We made extensive tests to verify
the reliability of this detection. We verified that the feature is
present both in the Polyn and in the ABBA reductions, as shown in
Fig.\ref{spectrum}.  We then inspected all the 75 individual spectra
to ensure that the feature is not due to an artifact, and that it is
still detected when we separately summed the data in two halves.
Because of the large color break ($z-Y>2.1$) measured in the HUDF data
and the non--detection in the $BVI$ bands, an identification of this line
with a lower redshift [OII] or H$\alpha$ would imply a very peculiar
SED, unlike that of currently known galaxies. This cannot be excluded a
priori.

We note that there is no evidence of the asymmetry
that is expected (but not required, see discussion below) for a
$z\simeq 7$ galaxy, although the S/N is too poor to reach any firm
conclusion about this.

Based on these tests, we conclude that the feature is likely real and
due to  \lya emission from a $z=6.972$ galaxy ($z=6.970$ if
computed at the blue edge of the line), although this should be
validated by independent and possibly deeper observations. 
No continuum is
detected in the spectrum: if we estimate it from the Hawk--I Y-band
magnitude (Table 1), the line flux translates into an observed EW of
103\AA, corresponding to $13$\AA\  if placed at $z=6.972$.

 We do not identify any other emission lines from objects in our sample.
 We only detect a faint continuum from two objects, namely G2\_2370 
 (the brightest in our sample) and the brown dwarf candidate of
 \citet{Mannucci2007}.  In both cases, the continuum is
 consistent with the broad--band magnitudes but the low S/N
 prevents us from deriving any robust information about their
 spectral type or redshift.

\section{The expected number of \lya detections}

The key result of our observations is the lack of prominent \lya
emission lines in our sample, which may imply a rapid evolution in the
physical properties of $z>6$ galaxies and/or in the surrounding
IGM. To quantify this issue, we have carried out the following Monte
Carlo simulations under the assumptions that {\it a)} all our 7
candidates are indeed $z\simeq 7$ galaxies; and {\it b)} the distribution
of the \lya intensity in galaxies as a function of their rest--frame
continuum magnitude $M_{UV}$ does not change significantly from
$z=4-6$ to $z=7$.

\begin{figure}
\epsscale{1.2}
\plotone{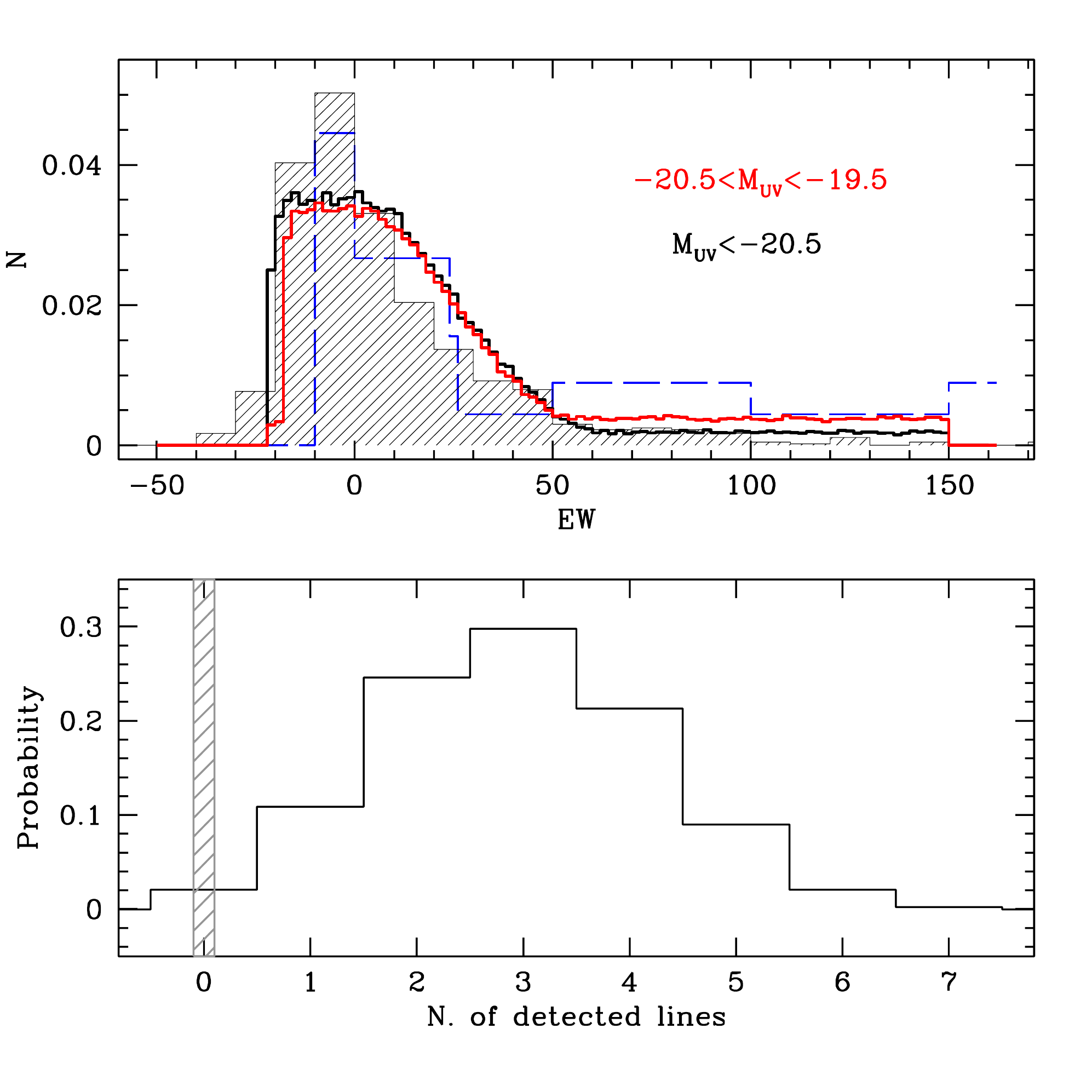}
\caption{ Simulations on the expected number of \lya emitters in our
   sample. {\it Upper:} The adopted distribution of rest-frame EW for
   the two extreme values of rest-frame UV luminosities in our sample
   (red and black continuous histogram). The shaded histogram  shows
   the \citet{Shapley2003} distribution at $z\simeq 3$ and the
   blue dashed histogram shows the \citet{Stanway2007} distribution at $z\simeq 4-6$.
   {\it Lower:} The resulting probability of detecting $N$ lines with
   S/N$>10$ in our sample, using the simulations described in the
   text. We observe no \lya with S/N$>10$.}.\label{EW_sim}
 \end{figure}

For the redshift distribution expected for our sample we use the
result by C10 (see their Fig 7), which has a broad maximum from
$z=6.4$ to $z=7.1$ and tails that extend to $z=6$ and $z=7.5$.  The
distribution of the \lya intensity in galaxies at $z=3-6$ has been
investigated in a number of studies (S07, V09, S10, D10), showing that the
intensity of \lya is anti--correlated with rest--frame UV luminosity.
No measure of the dependence of the EW distribution as a function of
$M_{UV}$ has been obtained, however.  We model the EW distribution
assuming that at EW$>0$ it is represented by a Gaussian centered on
EW=0 with an additional constant tail up to 150\AA, and at EW$<0$ by a
constant level down to some EW$_{min}$ value, and null below. We take
the width of the Gaussian and the two tails to reproduce the results
of V09 and S10 at different rest--frame magnitudes. Specifically, we
derive from the bright galaxies in V09 a standard deviation for the
Gaussian of 10\AA, and assume that it is constant at all magnitudes. We then
divide our sample in two luminosity bins ($-20.5< M_{UV}$ and
$-20.5<M_{UV}<-19.5$) and adjust the two tails in order to reproduce
the fraction of galaxies with $EW>50$\AA\ given by S10 and the
fraction of galaxies with EW~$>5$ and EW$>20$ \AA\ (for the two bins,
respectively), as given by the V09 data.  The resulting distributions
are shown in Figure \ref{EW_sim} for the two magnitude bins, and are
reasonably similar in shape to the EW distribution at $z\simeq
  5-6$ (S07), and show a moderate evolution from the  $z\simeq 3-5$
  \citep[][D10]{Shapley2003} one.

We then compute the probability of detecting $N$ \lya lines at a given
S/N in our sample of 7 objects. For each object we randomly extract a
redshift from the C10 distribution, we compute the corresponding
$M_{UV}$ from the observed $Y$ band magnitude (taking into account the
IGM absorption at that redshift), and we then randomly extract an EW
from the corresponding distribution. If the EW is larger than the
minimum detectable EW at the corresponding wavelength
(Fig.\ref{EW_lambda}) for a given S/N we conclude that the object
would be detected.  We assume FWHM=10\AA\ for the line, as found
  at $z=6.9$ by \citet{Iye2006} (see also Fig.\ref{EW_lambda}).
  Clearly, intrinsically broader lines would be harder to detect.  We
perform this exercise $10^5$ times over the whole sample, requiring
S/N$>10$ for the detection (larger than the S/N of the possible
detection in G2\_1408), and we finally obtain the probability
distribution shown in the lower panel of Fig.~\ref{EW_sim}. Under
these assumptions, the probability of detecting no \lya line in our
sample is very small, about 2\%, while the typical number of \lya that
we should have detected is between 2 and 4. We also find a low
probability (4\%) of having 1 detection at S/N$>5$, as found in our
sample.  The same probability adopting the S07 distribution would be
much smaller ($\simeq  10^{-3}$), because of the substantial
tail of objects with large EW.  Even using the \citet{Shapley2003}
distribution, which has a lower fraction of high EW objects, the
probability is still rather low (9\%). We conclude that, with all the
obvious caveats due to the small size of our sample and to possible
observational mishaps, the lack of prominent \lya lines in our sample
is statistically significant.

\section{Implications and Discussion}

On a practical level, our results show how challenging it is to obtain
large samples of spectroscopically confirmed galaxies at $z>6.5$ with
current instrumentation, especially if one aims at reaching the level
of completeness ($\gg 50\%$) needed to robustly measure the luminosity
function. Our observations imply that this goal will have to wait
until a future generation of instruments is available, either 8m
telescopes equipped with multi-object spectrographs more efficient in the $z$
and $Y$ bands or, more likely, the new generation of telescopes, such
as the {\it James Webb Space Telescope} or 20-40m ground-based facilities.
Nonetheless, our analysis appears to show that the failure to detect
prominent \lya in our sample is not only due to the insufficiency of 
current instrumentation.

 One possibility is that a significant fraction of the candidates are
 lower redshift interlopers.  We test this possibility by
 extrapolating to $z\sim 7$ the observed contamination in
 spectroscopic samples at $z\sim 4$, 5 and 6 (V09, Table 4 of B, V and
 i dropouts), which increases with redshift. We assume  that
 amongst our z--band dropout sample the fraction of contaminants could
 be $\sim$25\%, i.e., 2 out of 7 candidates.  This estimate may
   be pessimistic, given the excellent photometric quality of the
   Hawk--I and WFC3 data, and the more careful cleaning of lower $z$
   interlopers compared to the V09 samples. However, the contaminant
   population may be changing at higher redshifts, and different and
   previously unstudied galaxy types may be entering the selection
   window.  Ignoring these uncertainties, we repeated the Monte Carlo
 simulation for all possible choices of 5 candidates from our 7,
 finding that the probability of detecting no \lya line at S/N$>10$ is
 still rather low, being typically 8\%, and only in one case reaching
 15\% (this range depends on which candidates are excluded from the
 sample).

Another explanation for the paucity of \lya detections would be 
physical evolution of either the galaxies or the surrounding IGM.  The
intrinsic strength of the \lya emission is expected to increase at
higher redshift as
galaxies become more metal- and dust-poor. The probability that
these photons escape the galaxy and its surroundings, however, depends
on a series of complex (and not fully characterized) phenomena in the
IGM surrounding the galaxies \citep[][and references
therein]{Zheng2010}, including the relative geometry and dynamics of
gas and dust, e.g., backward scattering from wind--driven outflowing
shells (which can even boost the strength of the lines), or absorption
by the damping wings of in-falling IGM along the line of sight.  The
presence of HI in proximity to the source can result in an absorption
of the intrinsic \lya by one order of magnitude or
more \citep{Zheng2010}, along with a broadening and redshifting of the
emerging line profile. Thus, one explanation for the lack of
\lya detections in our sample is a significant increase in the HI
fraction of the IGM, $\chi_{HI}$, at $z\sim 7$, leading to a stronger
absorption of the \lya flux.  A similar effect could explain the
observed decrease in the number of \lya emitters at $z\simeq 7$
\citep[][Cl\'ement et al., in prep.]{Ota2010}, but results from these
surveys are still contradictory \citep{Ouchi2010, Hu2010}. Detailed
simulations \citep{Dayal2010b} show that the IGM absorption increases
dramatically when the Universe is not fully ionized, leading to a
significant absorption of the emerging \Lya. The timescale of this
effect around star--forming galaxies is of the order of 100 Myr
\citep{Dayal2010b}, shorter than the interval of cosmic time between
$z\simeq 6.$ and $z\simeq 7$. An additional prediction is that the
asymmetry in the line profile is smoothed by the velocity structure of
the infalling IGM \citep{Dayal2008}. Unfortunately, the modest S/N in
our only detection is too low to address this effect quantitatively.

In conclusion, this work shows that the spectroscopic confirmation of
$z\simeq 7$ galaxy candidates is a challenging effort. However, these
difficulties may not be only due to our current technological limits,
but may also reflect the long--sought first evidence of
the reionization process in the early Universe. Future surveys will
definitely solve this fascinating puzzle.

\acknowledgments Observations were carried out using the Very Large
Telescope at the ESO Paranal Observatory under Programme IDs
084.A-095, 181.A-0717. We thank an anonymous referee for precious
comments. We are grateful to P. M\o ller and the whole ESO staff for
their assistance during the execution of service observations. We
acknowledge partial financial support by ASI.


\begin{thebibliography}{34}
\expandafter\ifx\csname natexlab\endcsname\relax\def\natexlab#1{#1}\fi

\bibitem[{{Bouwens} {et~al.}(2010{\natexlab{a}}){Bouwens}, {Illingworth},
  {Oesch}, {Stiavelli}, {van Dokkum}, {Trenti}, {Magee}, {Labb{\'e}}, {Franx},
  {Carollo}, \& {Gonzalez}}]{Bouwens2009b}
{Bouwens}, R.~J., {Illingworth}, G.~D., {Oesch}, P.~A., {et~al.}
  2010{\natexlab{a}}, \apjl, 709, L133

\bibitem[{{Bouwens} {et~al.}(2010{\natexlab{b}}){Bouwens}, {Illingworth},
  {Oesch}, {Trenti}, {Stiavelli}, {Carollo}, {Franx}, {van Dokkum},
  {Labb{\'e}}, \& {Magee}}]{Bouwens2010b}
{Bouwens}, R.~J., {Illingworth}, G.~D., {Oesch}, P.~A., {et~al.}
  2010{\natexlab{b}}, \apjl, 708, L69

\bibitem[{{Bouwens} {et~al.}(2004){Bouwens}, {Thompson}, {Illingworth},
  {Franx}, {van Dokkum}, {Fan}, {Dickinson}, {Eisenstein}, \&
  {Rieke}}]{Bouwens2004}
{Bouwens}, R.~J., {Thompson}, R.~I., {Illingworth}, G.~D., {et~al.} 2004,
  \apjl, 616, L79

\bibitem[{{Bunker} {et~al.}(2009){Bunker}, {Wilkins}, {Ellis}, {Stark},
  {Lorenzoni}, {Chiu}, \& {Lacy}}]{Bunker2009}
{Bunker}, A., {Wilkins}, S., {Ellis}, R., {et~al.} 2009, ArXiv e-prints

\bibitem[{{Castellano} {et~al.}(2010){Castellano}, {Fontana}, {Boutsia},
  {Grazian}, {Pentericci}, {Bouwens}, {Dickinson}, {Giavalisco}, {Santini},
  {Cristiani}, {Fiore}, {Gallozzi}, {Giallongo}, {Maiolino}, {Mannucci},
  {Menci}, {Moorwood}, {Nonino}, {Paris}, {Renzini}, {Rosati}, {Salimbeni},
  {Testa}, \& {Vanzella}}]{Castellano2010}
{Castellano}, M., {Fontana}, A., {Boutsia}, K., {et~al.} 2010, \aap,
511, A20+ (C10)

\bibitem[{{Dayal} {et~al.}(2008){Dayal}, {Ferrara}, \& {Gallerani}}]{Dayal2008}
{Dayal}, P., {Ferrara}, A., \& {Gallerani}, S. 2008, \mnras, 389, 1683

\bibitem[{{Dayal} {et~al.}(2010){Dayal}, {Maselli}, \& {Ferrara}}]{Dayal2010b}
{Dayal}, P., {Maselli}, A., \& {Ferrara}, A. 2010, ArXiv e-prints

\bibitem[{{Douglas} {et~al.}(2010){Douglas}, {Bremer}, {Lehnert}, {Stanway}, \&
  {Milvang-Jensen}}]{Douglas2010}
{Douglas}, L.~S., {Bremer}, M.~N., {Lehnert}, M.~D., {Stanway}, E.~R., \&
  {Milvang-Jensen}, B. 2010, ArXiv e-prints (D10)

\bibitem[{{Finkelstein} {et~al.}(2009){Finkelstein}, {Papovich}, {Giavalisco},
  {Reddy}, {Ferguson}, {Koekemoer}, \& {Dickinson}}]{Finkelstein2009}
{Finkelstein}, S.~L., {Papovich}, C., {Giavalisco}, M., {et~al.} 2009, ArXiv
  e-prints

\bibitem[{{Giavalisco}(2002)}]{Giavalisco2002}
{Giavalisco}, M. 2002, \araa, 40, 579

\bibitem[{{Giavalisco} {et~al.}(1996){Giavalisco}, {Koratkar}, \&
  {Calzetti}}]{Giavalisco1996}
{Giavalisco}, M., {Koratkar}, A., \& {Calzetti}, D. 1996, \apj, 466, 831

\bibitem[{{Greiner} {et~al.}(2009){Greiner}, {Kr{\"u}hler}, {Fynbo}, {Rossi},
  {Schwarz}, {Klose}, {Savaglio}, {Tanvir}, {McBreen}, {Totani}, {Zhang}, {Wu},
  {Watson}, {Barthelmy}, {Beardmore}, {Ferrero}, {Gehrels}, {Kann}, {Kawai},
  {Yolda{\c s}}, {M{\'e}sz{\'a}ros}, {Milvang-Jensen}, {Oates}, {Pierini},
  {Schady}, {Toma}, {Vreeswijk}, {Yolda{\c s}}, {Zhang}, {Afonso}, {Aoki},
  {Burrows}, {Clemens}, {Filgas}, {Haiman}, {Hartmann}, {Hasinger}, {Hjorth},
  {Jehin}, {Levan}, {Liang}, {Malesani}, {Pyo}, {Schulze}, {Szokoly}, {Terada},
  \& {Wiersema}}]{Greiner2009}
{Greiner}, J., {Kr{\"u}hler}, T., {Fynbo}, J.~P.~U., {et~al.} 2009, \apj, 693,
  1610

\bibitem[{{Hickey} {et~al.}(2010){Hickey}, {Bunker}, {Jarvis}, {Chiu}, \&
  {Bonfield}}]{Hickey2009}
{Hickey}, S., {Bunker}, A., {Jarvis}, M.~J., {Chiu}, K., \& {Bonfield}, D.
  2010, \mnras, 404, 212

\bibitem[{{Hu} {et~al.}(2010){Hu}, {Cowie}, {Barger}, {Capak}, {Kakazu}, \&
  {Trouille}}]{Hu2010}
{Hu}, E.~M., {Cowie}, L.~L., {Barger}, A.~J., {et~al.} 2010, ArXiv e-prints

\bibitem[{{Iye} {et~al.}(2006){Iye}, {Ota}, {Kashikawa}, {Furusawa},
  {Hashimoto}, {Hattori}, {Matsuda}, {Morokuma}, {Ouchi}, \&
  {Shimasaku}}]{Iye2006}
{Iye}, M., {Ota}, K., {Kashikawa}, N., {et~al.} 2006, \nat, 443, 186

\bibitem[{{Labb{\'e}} {et~al.}(2010){Labb{\'e}}, {Gonz{\'a}lez}, {Bouwens},
  {Illingworth}, {Oesch}, {van Dokkum}, {Carollo}, {Franx}, {Stiavelli},
  {Trenti}, {Magee}, \& {Kriek}}]{Labbe2010}
{Labb{\'e}}, I., {Gonz{\'a}lez}, V., {Bouwens}, R.~J., {et~al.} 2010, \apjl,
  708, L26

\bibitem[{{Mannucci} {et~al.}(2007){Mannucci}, {Buttery}, {Maiolino},
  {Marconi}, \& {Pozzetti}}]{Mannucci2007}
{Mannucci}, F., {Buttery}, H., {Maiolino}, R., {Marconi}, A., \& {Pozzetti}, L.
  2007, \aap, 461, 423

\bibitem[{{McLure} {et~al.}(2010){McLure}, {Dunlop}, {Cirasuolo}, {Koekemoer},
  {Sabbi}, {Stark}, {Targett}, \& {Ellis}}]{Mclure2009b}
{McLure}, R.~J., {Dunlop}, J.~S., {Cirasuolo}, M., {et~al.} 2010, \mnras, 403,
  960

\bibitem[{{Oesch} {et~al.}(2010){Oesch}, {Bouwens}, {Illingworth}, {Carollo},
  {Franx}, {Labb{\'e}}, {Magee}, {Stiavelli}, {Trenti}, \& {van
  Dokkum}}]{Oesch2009b}
{Oesch}, P.~A., {Bouwens}, R.~J., {Illingworth}, G.~D., {et~al.} 2010, \apjl,
  709, L16

\bibitem[{{Ota} {et~al.}(2010){Ota}, {Iye}, {Kashikawa}, {Shimasaku}, {Ouchi},
  {Totani}, {Kobayashi}, {Nagashima}, {Harayama}, {Kodaka}, {Morokuma},
  {Furusawa}, {Tajitsu}, \& {Hattori}}]{Ota2010}
{Ota}, K., {Iye}, M., {Kashikawa}, N., {et~al.} 2010, ArXiv e-prints

\bibitem[{{Ouchi} {et~al.}(2009){Ouchi}, {Mobasher}, {Shimasaku}, {Ferguson},
  {Fall}, {Ono}, {Kashikawa}, {Morokuma}, {Nakajima}, {Okamura}, {Dickinson},
  {Giavalisco}, \& {Ohta}}]{Ouchi2009}
{Ouchi}, M., {Mobasher}, B., {Shimasaku}, K., {et~al.} 2009, \apj, 706, 1136

\bibitem[{{Ouchi} {et~al.}(2010){Ouchi}, {Shimasaku}, {Furusawa}, {SAITO},
  {Yoshida}, {Akiyama}, {Ono}, {Yamada}, {Ota}, {Kashikawa}, {Iye}, {Kodama},
  {Okamura}, {Simpson}, \& {Yoshida}}]{Ouchi2010}
{Ouchi}, M., {Shimasaku}, K., {Furusawa}, H., {et~al.} 2010, ArXiv e-prints

\bibitem[{{Pentericci} {et~al.}(2010){Pentericci}, {Grazian}, {Scarlata},
  {Fontana}, {Castellano}, {Giallongo}, \& {Vanzella}}]{Pentericci2010}
{Pentericci}, L., {Grazian}, A., {Scarlata}, C., {et~al.} 2010, \aap, 514, A64+

\bibitem[{{Reddy} \& {Steidel}(2009)}]{Reddy2009}
{Reddy}, N.~A. \& {Steidel}, C.~C. 2009, \apj, 692, 778

\bibitem[{{Salvaterra} {et~al.}(2009){Salvaterra}, {Della Valle}, {Campana},
  {Chincarini}, {Covino}, {D'Avanzo}, {Fern{\'a}ndez-Soto}, {Guidorzi},
  {Mannucci}, {Margutti}, {Th{\"o}ne}, {Antonelli}, {Barthelmy}, {de Pasquale},
  {D'Elia}, {Fiore}, {Fugazza}, {Hunt}, {Maiorano}, {Marinoni}, {Marshall},
  {Molinari}, {Nousek}, {Pian}, {Racusin}, {Stella}, {Amati}, {Andreuzzi},
  {Cusumano}, {Fenimore}, {Ferrero}, {Giommi}, {Guetta}, {Holland}, {Hurley},
  {Israel}, {Mao}, {Markwardt}, {Masetti}, {Pagani}, {Palazzi}, {Palmer},
  {Piranomonte}, {Tagliaferri}, \& {Testa}}]{Salvaterra2009}
{Salvaterra}, R., {Della Valle}, M., {Campana}, S., {et~al.} 2009, \nat, 461,
  1258

\bibitem[{{Schaerer} \& {de Barros}(2010)}]{Schaerer2010}
{Schaerer}, D. \& {de Barros}, S. 2010, ArXiv e-prints

\bibitem[{{Shapley} {et~al.}(2003){Shapley}, {Steidel}, {Pettini}, \&
  {Adelberger}}]{Shapley2003}
{Shapley}, A.~E., {Steidel}, C.~C., {Pettini}, M., \& {Adelberger}, K.~L. 2003,
  \apj, 588, 65 (S03)

\bibitem[{{Stanway} {et~al.}(2007){Stanway}, {Bunker}, {Glazebrook}, {Abraham},
  {Rhoads}, {Malhotra}, {Crampton}, {Colless}, \& {Chiu}}]{Stanway2007}
{Stanway}, E.~R., {Bunker}, A.~J., {Glazebrook}, K., {et~al.} 2007, \mnras,
  376, 727 (S07)

\bibitem[{{Stark} {et~al.}(2010){Stark}, {Ellis}, {Chiu}, {Ouchi}, \&
  {Bunker}}]{Stark2010}
{Stark}, D.~P., {Ellis}, R.~S., {Chiu}, K., {Ouchi}, M., \& {Bunker}, A. 2010,
  ArXiv e-prints

\bibitem[{{Tanvir} {et~al.}(2009){Tanvir}, {Fox}, {Levan}, {Berger},
  {Wiersema}, {Fynbo}, {Cucchiara}, {Kr{\"u}hler}, {Gehrels}, {Bloom},
  {Greiner}, {Evans}, {Rol}, {Olivares}, {Hjorth}, {Jakobsson}, {Farihi},
  {Willingale}, {Starling}, {Cenko}, {Perley}, {Maund}, {Duke}, {Wijers},
  {Adamson}, {Allan}, {Bremer}, {Burrows}, {Castro-Tirado}, {Cavanagh}, {de
  Ugarte Postigo}, {Dopita}, {Fatkhullin}, {Fruchter}, {Foley}, {Gorosabel},
  {Kennea}, {Kerr}, {Klose}, {Krimm}, {Komarova}, {Kulkarni}, {Moskvitin},
  {Mundell}, {Naylor}, {Page}, {Penprase}, {Perri}, {Podsiadlowski}, {Roth},
  {Rutledge}, {Sakamoto}, {Schady}, {Schmidt}, {Soderberg}, {Sollerman},
  {Stephens}, {Stratta}, {Ukwatta}, {Watson}, {Westra}, {Wold}, \&
  {Wolf}}]{Tanvir2009}
{Tanvir}, N.~R., {Fox}, D.~B., {Levan}, A.~J., {et~al.} 2009, \nat, 461, 1254

\bibitem[{{Vanzella} {et~al.}(2009){Vanzella}, {Giavalisco}, {Dickinson},
  {Cristiani}, {Nonino}, {Kuntschner}, {Popesso}, {Rosati}, {Renzini}, {Stern},
  {Cesarsky}, {Ferguson}, \& {Fosbury}}]{Vanzella2009}
{Vanzella}, E., {Giavalisco}, M., {Dickinson}, M., {et~al.} 2009, \apj, 695,
  1163 (V09)

\bibitem[{{Wilkins} {et~al.}(2010{\natexlab{a}}){Wilkins}, {Bunker}, {Ellis},
  {Stark}, {Stanway}, {Chiu}, {Lorenzoni}, \& {Jarvis}}]{Wilkins2009}
{Wilkins}, S.~M., {Bunker}, A.~J., {Ellis}, R.~S., {et~al.} 2010{\natexlab{a}},
  \mnras, 403, 938

\bibitem[{{Wilkins} {et~al.}(2010{\natexlab{b}}){Wilkins}, {Bunker},
  {Lorenzoni}, \& {Caruana}}]{Wilkins2010}
{Wilkins}, S.~M., {Bunker}, A.~J., {Lorenzoni}, S., \& {Caruana}, J.
  2010{\natexlab{b}}, ArXiv e-prints

\bibitem[{{Zheng} {et~al.}(2010){Zheng}, {Cen}, {Trac}, \&
  {Miralda-Escud{\'e}}}]{Zheng2010}
{Zheng}, Z., {Cen}, R., {Trac}, H., \& {Miralda-Escud{\'e}}, J. 2010, \apj,
  716, 574

\end{thebibliography}
\end{document}